# Self Excitation of Nano-Mechanical Pillars

Hyun S. Kim[*], Hua Qin[*], and Robert H. Blick

*Electrical & Computer Engineering*

*University of Wisconsin-Madison*

*1415 Engineering Drive, Madison, WI 53706-1691, USA.*

8/13/2007 0:52:45 a8/p8

**Self excitation is a mechanism which is ubiquitous for electromechanical power devices such as electrical generators. This is conventionally achieved by making use of the magnetic field component in electrical generators [1], where a good example are the overall visible wind farm turbines [2]. In other words, a static force, like wind acting on the rotor blades, can generate a resonant excitation at a certain mechanical frequency. For nanomechanical systems [3,4,5] such a self excitation (SE) mechanism is highly desirable as well, since it can generate mechanical oscillations at radio frequencies by simply applying a DC bias voltage. This is of great importance for low-power signal communication devices and detectors, as well as for mechanical computing elements. For a particular nanomechanical system – the single electron shuttle – this effect was predicted some time ago by Gorelik *et al*. [6]. Here, we use a nano-electromechanical single electron transistor (NEMSET) to demonstrate self excitation for both the soft and hard regime, respectively. The ability to use self excitation in nanomechanical systems may enable the detection of quantum mechanical backaction effects [7] in direct tunneling, macroscopic quantum tunneling [8], and rectification [9]. All these effects have so far been over shadowed by the large driving voltages, which had to be applied.**

Straight forward shuttling in semiconductor nanostructures of electrons was demonstrated in different configurations of the NEMSET [10,11]. The main limitation in resolving quantum mechanical effects, such as Coulomb blockade (CB) of electron transport is given by the excitation mechanism: an AC driver signal applied to the source electrode leads via the resonant Coulomb force (RCF) mechanism [12,13] to the onset of mechanical resonance at the eigenfrequency of a NEMSET. However, the signal level commonly required is of the order of several volts, thus overshadowing CB effects. For an estimate for the Coulomb energy we can use the orthodox CB-model, which relates the total capacitance of the pillar $C_\Sigma = C_s + C_d$ to the charging energy $E_C = e^2/2C_\Sigma$. This in turn gives an energy of about 30 meV for a small metallic NEMSET of some 10 nm diameter (with a typical capacitance of some 10 aF). This is just sufficient to reveal CB at room temperature, however, it has to be combined with a proper excitation mechanism for mechanical resonances, such as self excitation (SE).

The current generation of NEMSETs are processed as nanopillars and placed in a microwave coplanar wave guide between two metallic contacts in silicon-on-insulator wafer material. On top of the 50 nm wide and 300 nm tall pillars a thin layer of gold is deposited, functioning as the electron island (see Fig. 1). In contrast to earlier work [11,14], the nanopillars we are investigating here possess resonance frequencies at the lower end of the radio frequency spectrum. This is due to the more 'mushroom'-like shape of the pillars, as can be seen in Fig. 1(a). We anticipated that with such a shape the regime of self-excitation can be established more easily.

In order to test both regimes of self excitation we fabricated nanopillars in different gate geometries (termed *flat* and *sharp* in the following), which translates into soft SE and hard SE, as outlined in the theory by Gorelik *et al.* [6]. In essence it is a competition between the electric field energy (~ $\omega = 2\pi f$ and a DC bias) and the mechanical energy (~ $\omega_m = 2\pi f_m$). With altering the gate geometry we have chosen to primarily vary effective electric field applied to the nanopillar, while using pillars of similar shape.

In the following we will first focus on the analysis of soft-SE where the nanopillar is placed between two flat electrodes, as shown in Fig. 1(a). In order to have a good understanding of the effect of a single nanopillar on the transmission through such a contact, we also fabricated a second sample without a pillar (see Fig. 1(b)), termed 'gap'. The samples are placed in a probe station with high-frequency probe tips (DC – 50 GHz). The probe station is evacuated to a pressure of $10^{-4}$ mbar and all measurements are performed at room temperature. Both devices are placed in a coplanar wave guide for probing the electromechanical response vs. radio frequency (RF) applied (see Fig. 1(c)). Commonly an RF synthesizer is combined with a DC bias ($V_{DC}$) via a bias-tee to drive the sample capacitively with a voltage $V_{ac}$ at frequency $\omega_0$. This induces a mechanical displacement via resonant Coulomb force excitation [12,13], if the RF signal matches a mechanical eigenfrequency of the nanopillar $\omega_m$. Depending

*Kim, Qin & Blick*                                                                                                          4

on the amplitudes of $V_{ac}$ and the bias voltage $V_{dc}$, a current $I_D$ of some 10 pA to 10 nA is shuttled from drain to source. Finally, the current is amplified with a standard current amplifier.

For determining the mechanical self excitation we made use of the intrinsic signal mixing properties of the NEMSET, i.e. the nonlinear characteristic: conventionally two phase-locked synthesizers with a combined output signal $\{V_{ac}^0(\omega_0) + V_{ac}^1(\omega_1)\}$ are applied to the input port of the sample with a nonlinear *IV*-characteristic [14,15]. The difference frequency defined by $\delta\omega = |\omega_1 - \omega_0|$ or multiples of it are the mixing product, which can be used as a reference for the lock-in amplifier (LIA). Here, we do not focus onto the signal mixing process itself (which we describe in detail elsewhere [15]), but make use of the *same principle for probing self-excitation*: we apply one electromagnetic signal at $\omega_0$, but then use the mechanical frequency $\omega_m$ of the nanopillar (self-excited via the DC bias voltage) as the second component of the mixing signal. It has to noted that the electromagnetic RF-signal is applied at sub-threshold and only triggers the mixing process. The finally resulting signal can then be traced with the lock-in amplifier at the reference frequency $\Delta\omega = \omega_0 - \omega_m$. This approach ensures that we can accurately monitor self-excitation, as will be shown below.

The single nanopillar is placed in a coplanar waveguide (CPW) for improved impedance and hence power matching. Transport of electrons occurs by shuttling via the nanopillar, that is electrons tunnel onto the metallic island on top of the pillar when it is close to the source contact and then leave the pillar once it is close to the drain electrode. In addition to conventional tunneling this electron shuttling mechanism is supported by thermionic emission and field emission. The effective field enhancement supporting field emission via the electron shuttle is documented in the data for the *flat* electrodes shown in Fig. 2: the *IV*-characteristic of the nanopillar (no RF applied) is compared to the characteristic of a CPW-junction *without* a nanopillar. Obviously, electron transport has a strongly reduced threshold with the nanopillar in place by a factor of five. This is clear evidence for the control of electron transport by the nanopillar. The detailed description of field emission and the mechanical resonance spectra is given elsewhere [13,14]. Clearly visible is also the strongly nonlinear *IV*-characteristic. The dominance of field emission is underlined in Fig. 2(b), where the traces from (a) are plotted in the standard Fowler-Nordheim fashion. As expected field emission is strongest for large bias voltages $V_{bias}$, while for small bias conventional tunneling through the nanopillar prevails. From the extrapolation of the solid lines we obtain an order of magnitude difference between the pure gap (~ 40V) and the nanopillar junction (~ 4V) for the onset for field emission.

The next step in the measurements is to probe the current spectra of the nanopillar: in Fig. 3(a) the response of the nanomechanical resonator to an AC-excitation is shown in the direct current $I_D$. An AC signal of 19 dBm is superimposed to the DC bias varied from 0 to 16V. This is the standard mechanism for excitation of nanopillars based on the RCF method. The resulting apparent resonance at $\omega_m/2\pi$ = 10.5 MHz is fairly broad with a mechanical quality factor around $Q$ ~ 2.5. Per cycle of mechanical motion of the nanopillar it shuttles on average <$n$> = <$I$> / $2ef$ electrons, where $e$ is the elementary charge. For a current of 350 pA at the resonance of ~ 10 MHz, shown in Fig. 3(a), we obtain an average number of <$n$> ~ 100 electrons per cycle. Hence, the mixing signal is determined by single electrons being shuttled from source to drain.

Having shown that the *IV*-characteristic is nonlinear and that the nanopillar can be excited by an electromagnetic AC signal, we can now apply signal mixing. It is important to note, that we probe self excitation, i.e. instead of two electromagnetic signals, we apply a single *sub-threshold* AC signal ($\omega$) and mix this *with the mechanical eigenfrequency of the nanopillar at* $\omega_m$. Phase sensitive detection of a resulting signal at the offset frequency $\Delta\omega$ implies that the mechanical mode is indeed induced by the DC bias and hence is evidence for self excitation. The first step in this direction is shown in Fig. 3(b), where we compare a pure DC bias (black trace) with a superimposed {$V_{dc}$ + $V_{ac}$}. It is important to note that the effective signal strength of $V_{ac}$ now is below the threshold for RCF-excitation of a mechanical resonance (see Fig. 3 for details).

It appears clear that we supply the necessary energy for self excitation by the DC electric field. However, one could still argue that a rectification effect is predominant. Hence, only probing the highly sensitive mixing signal given as $G_{mix}$ ~ $d^2I/dV^2\ cos(\Delta\omega t)$ will provide the necessary information that it is indeed a mechanical displacement induced by a dominant DC voltage. In Fig. 3(c) this mixing signal is plotted vs. the probing AC frequency with the different traces corresponding to DC-bias values (traces are offset for clarity). As pointed out before these traces are obtained from phase sensitively amplifying the lock-in signal at $\Delta\omega$ under the condition $|V_{ac}| < |V_{th}|$ (see Fig. 3(b)). The mechanical Q-factor obtained from these measurements appears to be of the order of ~7, better than the one found in the direct current. This is most likely related to the fact that the mixing signal $G_{mix}$ is directly proportional to the displacement <x>, but neglects field emitted electrons, which are not directly associated with a mechanical motion. In other words the direct current determined in Fig. 3(a) shows shuttling of electrons, but also a contribution from field emitted electrons (as in standard RCF-excitation). Using this more sensitive mixing technique which relies on mechanical displacement of the nanopillar, we can now directly state the relation $G_{mix}$ ~ <x>. This effect corresponds to self-excitation of the nanopillar in the soft limit [6].

In Fig. 3(d) the detailed bias dependence of the self-excitation signal $G_{mix}$ under increasing AC-signal power is plotted. As predicted by Isacsson *et al.* [16] the traces follow a $V_{bias}^{1/2}$-dependence from the sharp onset of mechanical motion on. This is indicated by the fitted dashed lines. The threshold voltage is marked by $V_{th}$ with the onset at $V_{bias}$ ~ 4 V. Obviously, the onset can be clearly identified in contrast to the direct current recording in Fig. 3(b). This imprecision of the measurement in $I_D$ is due to the fact that the mixing signal truly relies on mechanical displacement, while the direct current contains contributions from field emitted electrons.

As predicted by Isacsson *et al*. [16] there exist two cases of self-excitation: the hard and the soft limit. Basically, the hard limit reveals a bistable state, which appears as a hysteresis in the *IV*-characteristic. The distinction between the hard and soft limit can be analyzed by comparing the electronic energy which is fed into the nanopillar by a constant bias voltage, while the nanopillar is oscillating at the mechanical energy ($\sim \omega_m$). In order to address the hard SE limit we altered the initially used electrode geometry by choosing a *sharp* electrode as source contact – see Fig. 4(a). Again a coplanar wave guide is used with the center lead being the signal line. We then placed two nanopillars – diameters of 60 nm (left) and 30 nm (right) – close to the signal line (see lower left inset). For the measurements we have chosen the left nanopillar, since the diameter is comparable to the ones used to study the soft limit of SE.

The competition between the electronic and the mechanical degrees of freedom then leads to a single stable mechanical resonance for the soft limit (flat electrode geometry) and a bistable state for the hard limit (sharp electrode geometry). The lower right hand side inset in Fig. 4(a) gives a numerical simulation of the electric field distribution between the two electrodes and the nanopillar. The color coding shows the field strength, as expected the highest field intensity is found between the sharp electrode and the pillar (red), while lowest intensities are found off the central symmetry axis at $y = 0$. As seen in the scanning electron microscope picture we can assume a slight misalignment of the nanopillar. The total energy of the nanopillar in such an electric field is sketched in the upper right hand side inset: the pure mechanical elastic energy will follow Hooke's law with $E_m = \kappa r^2$ with $\kappa$ being the restoring spring constant and $x^2 + y^2 = r^2$ the actual displacement from equilibrium (dashed black line). The sharp electrode now leads to an electrical field energy that is not symmetrical as compared to the flat electrode geometry: it shows a spike with increasing voltage (dash-dotted blue line). The total electromechanical energy is then given by the superposition of both contributions (solid red line). As seen a slight misalignment of the nanopillar with respect to the electrodes is sufficient to lead to an asymmetry between the two stable states α and β of the system. More precisely, α and β correspond to the two different mechanical modes.

In Fig. 4(b) and (c) two consecutive bias sweeps on the nanopillar are shown. The red trace in (b) was the initial measurement, while the blue trace (c) is taken after several sweeps. As seen we find the predicted hysteresis in both measurements. The width of the hysteresis is a direct measure of the energy difference between potential wells α and β which is of the order of 2eV (arrows in Fig. 4(c)). Such a bistability can be applied for memory applications or for probing entanglement in such a nanomechanical system, as suggested by Savelev *et al*. [8]. For reaching the quantum limit the device has to be cooled to mK-temperatures and the mechanical eigenfrequency has to be increased to the GHz-regime, which appears to be possible [17].

In summary we have demonstrated how to probe self excitation in a nanomechanical system by using mechanical mixing. This has apparent applications as a mechanical mixer for communication electronics, since the realization of a DC-driven mechanical resonance renders an external oscillator obsolete. In addition the device is intrinsically more sensitive, which will enable testing quantum fluctuations with great accuracy: under the assumption that the nanopillar is perfectly symmetric and placed in the center of the DC electric field, the pillar would remain in a stable state. A single fluctuation will then lead to the onset of mechanical oscillations in the electric field.

*Acknowledgements* – We like to thank the Air Force Office of Scientific Research (AFOSR – Grant No. F49620-03-1-0420), the Wisconsin Alumni Research Foundation, the National Science Foundation (MRSEC-IRG1) for support, and DARPA (NEMS project).

[*]: the first and second author contributed equivalently to this work.

# Figure Captions

**Figure 1.** (a) Scanning electron microscope picture of the nanopillar between two electrodes. The diameter of the pillar is 60 nm with a height of 250 nm – on the top a 45 nm gold layer is added. (b) Two-electrode configuration without a nanomechanical structure forming a simple gap for control measurements. The gap width is 110 nm. Note that the electrodes for both samples are forming *flat* capacitor plates. (c) Coplanar waveguide with nanopillar in the center and measurement circuit: at the source S, a signal generator provides the AC voltage $V_{AC}(t)$ at an incident power $P$ and at frequency $\omega$ with a superimposed DC bias. The reference signal $\Delta\omega = |\omega - \omega_m|$ is generated by mixing the incident electromagnetic signal at $\omega$ with the mechanical eigenfrequency $\omega_m$ of the nanopillar. Self-excitation is characterized by detecting the Lock-in amplifier (LIA) signal and the net current at drain D with a current amplifier.

**Figure 2.** (a) The direct current through the junction vs. bias voltage $V_{bias}$, while in (b) Fowler-Nordheim plots with the nanopillar and without (gap) are given. The onset of current flow for the nanopillar appears an order of magnitude below the one for the gap, as expected. Both plots indicate that the nanopillar strongly distorts the electromagnetic field and enhances the effective field strength purely by geometry.

**Figure 3.** Mechanical self-excitation of the nanopillar in the soft limit: (a) Direct current $I_D$ at mechanical resonance of $\omega_m/2\pi = f_m = 10.5$ MHz under conventional AC drive with a large amplitude of $V_{ac}(\omega)$. The different traces indicate the increase of the DC-bias leading to a rising resonance peak and an increase in background current. (b) Full bias dependence of the direct current at mechanical resonance measured at 300 K and $10^{-4}$ mbar. Comparison of DC-bias only ($V_{ac}$) and with a sub-threshold AC-signal ($V_{ac}$). (c) Mixing signal obtained with this AC-signal (at $\omega$ and DC-bias increased – here, the traces are offset for clarity). The mixing signal is directly proportional to the mechanical displacement and is only detected with the lock-in when the difference frequency of the electrical sub-threshold signal and the mechanical frequency match:

$|\omega - \omega_\mathrm{m}| = \Delta\omega$. (d) Bias dependence of the self-excitation signal with a power dependent voltage relation according to $V_\mathrm{bias}^{1/2}$, as predicted by Iscasson *et al*. [16]. The threshold voltage is indicated by $V_\mathrm{th}$, the onset can be clearly distinguished in contrast to the direct current recording in (b). This is due to the fact that the mixing signal truly relies on mechanical displacement only, while the direct current contains contributions from field emitted electrons.

**Figure 4.** Mechanical self-excitation in the hard limit: (a) Coplanar wave guide with an alternative sample geometry. As the lower left inset shows an electrode geometry was chosen with a *sharpened* electrode in contrast to the *flat* electrodes used before. The nanopillar dimensions of the left pillar are similar to the sample used for probing self-excitation in the soft limit [see Fig. 1(a)]. The lower right hand side inset shows a numerical simulation of the electromagnetic field distribution between the two electrodes and the nanopillar. As found in the electron micrograph we assumed a slight misalignment of the pillar. The total energy of the nanopillar in such an electric field is sketched in the upper right hand side inset: the pure mechanical elastic energy will have a square dependence on the spatial coordinate $r^2 = x^2 + y^2$ (dashed black line). Increasing the electric field builds up the potential indicated by the dash-dotted blue line. The superposition of both yields the total energy (solid red line) revealing two potential minima α and β. In (b) and (c) two consecutive bias sweeps on the nanopillar are shown. The red trace in (b) was the initial measurement, while the blue trace (c) shows the hysteresis after several sweeps. The width of the hysteresis is a measure of the energy difference between potential wells α and β.

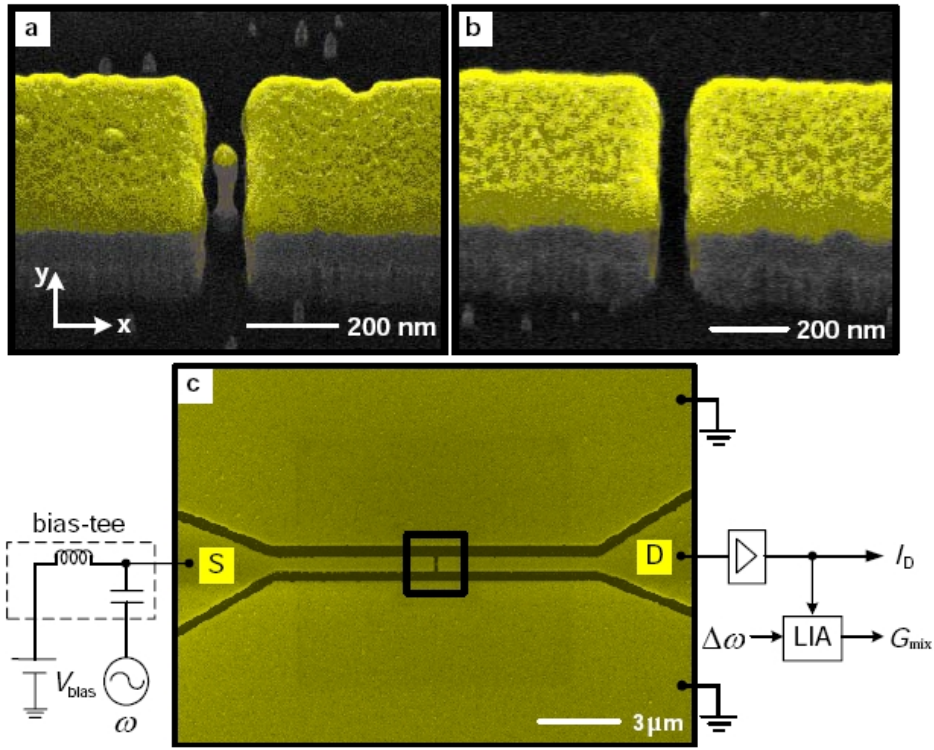

*Figure 1: Kim, Qin & Blick, 1/4*

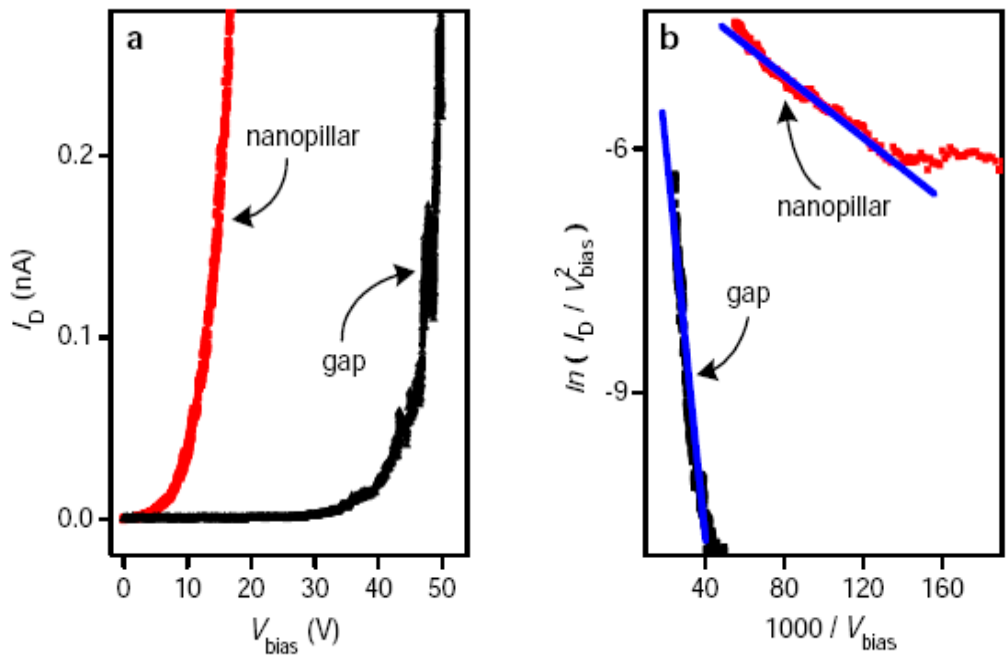

*Figure 2: Kim, Qin & Blick, 2/4*

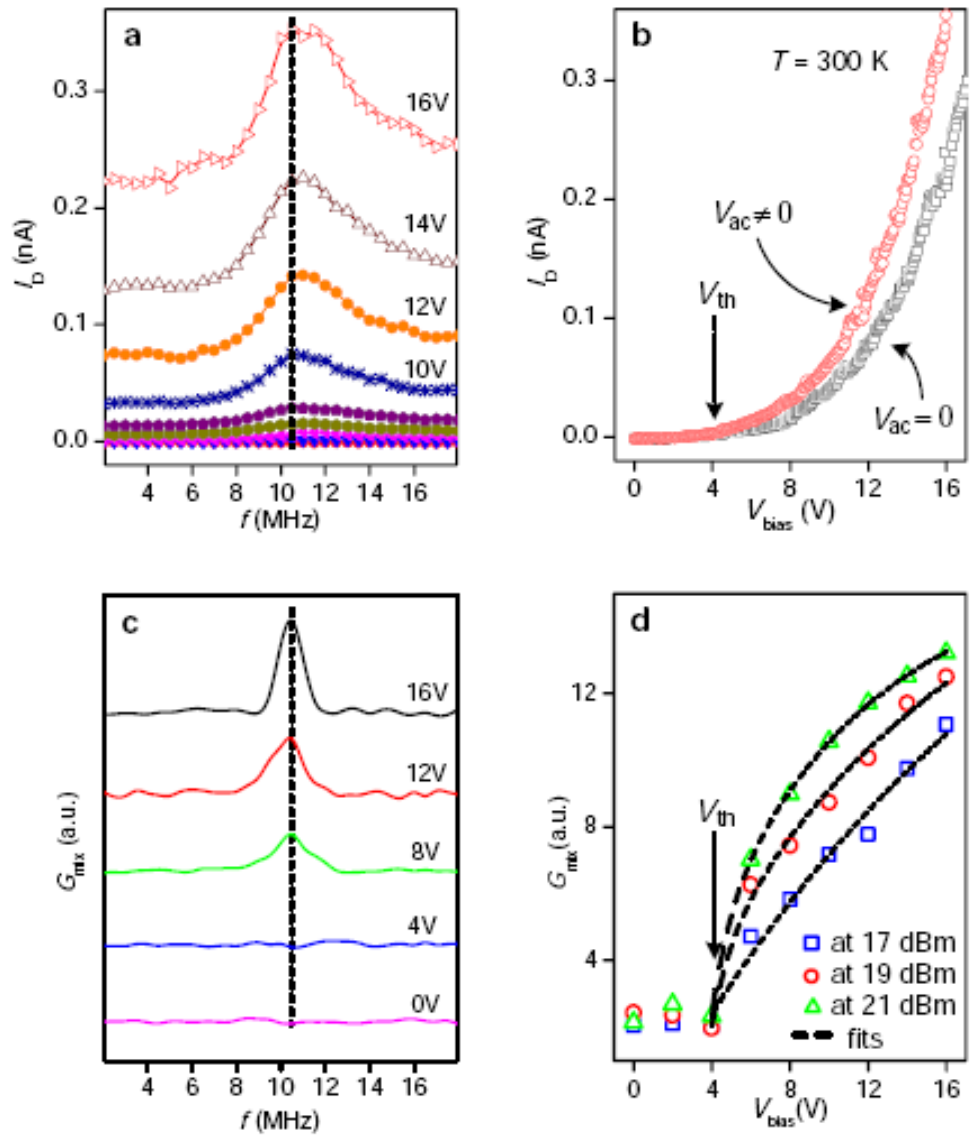

*Figure 3: Kim, Qin & Blick, 3/4*

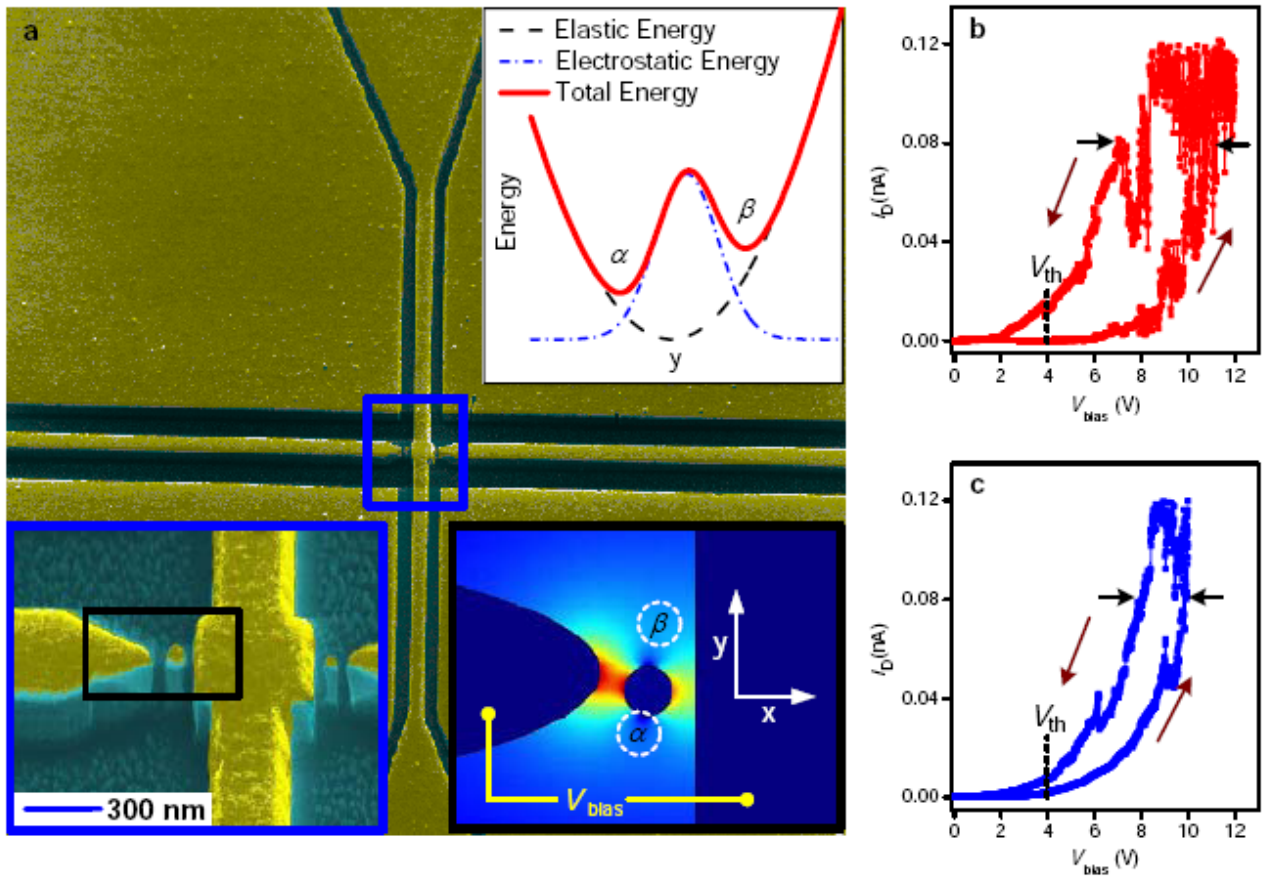

*Figure 4: Kim, Qin & Blick, 4/4*